\documentclass[aps,prl,twocolumn,superscriptaddress,amsmath]{revtex4-1}

\usepackage{graphicx}

\usepackage{xspace}
\usepackage{cleveref}
\usepackage{suffix}
\usepackage{ifthen}

\newcommand{\background}{}
\newcommand{\results}{}
\newcommand{\conclusion}{}
\WithSuffix\newcommand\author*[2]{\author{#1}}
\newenvironment{suppinfo}{}{}
\newcounter{suppinfo}
\newcommand{\sifile}[4][]{%
\refstepcounter{suppinfo}%
}
\crefname{suppinfo}{Supporting Information File}{Supporting
Information Files}
\crefformat{suppinfo}{Supporting Information File~#2#1#3}
\Crefformat{suppinfo}{Supporting Information File~#2#1#3}
\Crefname{figure}{Figure}{Figures}
\crefname{figure}{Fig.}{Figs.}

\begin{document}


\title{Charge and spin transport in mesoscopic superconductors}
\author{M. J. Wolf}
\author{F. H\"ubler}
\author{S. Kolenda}
\author*{D. Beckmann}{detlef.beckmann@kit.edu}
\affiliation{Karlsruher Institut f\"ur Technologie (KIT), Institut f\"ur Nanotechnologie}

\begin{abstract}
\background Nonequilibrium charge transport in superconductors has been investigated intensely in the 1970s and 80s, mostly in the vicinity of the critical temperature. Much less attention has been focussed on low temperatures, and the role of the quasiparticle spin. 
\results We report here on nonlocal transport in superconductor hybrid structures at very low temperatures. By comparing the nonlocal conductance obtained using ferromagnetic and normal-metal detectors, we discriminate charge and spin degrees of freedom. We observe spin injection and long-range transport of pure, chargeless spin currents in the regime of large Zeeman splitting. We elucidate charge and spin tranport by comparison to theoretical models.
\conclusion The observed long-range chargeless spin transport opens a new path to manipulate and utilize the quasiparticle spin in superconductor nanostructures.
\end{abstract}

\maketitle

\section{Introduction}
The investigation of spin-polarized transport in hybrid structures was pioneered in the 1970s with the discovery of spin-dependent tunneling into thin-film superconductors with a large Zeeman splitting by Tedrow and Meservey \cite{tedrow1971,meservey1994}. While much of the related basic physics such as tunneling magnetoresistance (TMR) \cite{julliere1975} and non-equilibrium spin injection \cite{johnson1985} was observed subsequently, spin-polarized transport did not attract much attention until the discovery of the giant magnetoresistance (GMR) \cite{baibich1988,binasch1989,zutic2004} and its technical applications. 

In superconductors, electrons are bound in Cooper pairs, which usually have a singlet structure and therefore carry only charge but no spin. The quasiparticle excitations, however, may carry both charge and spin. Nonequilibrium charge transport in superconductors has been investigated intensely in the 1970s and 80s, mostly in the vicinity of the critical temperature \cite{clarke1972,tinkham1972,langenberg} and more recently also in the low-temperature regime \cite{yagi2006,huebler2010,arutyunov2011}. In contrast, only few experiments on quasiparticle spin transport \cite{johnson1994} have been reported, and the subject remains poorly understood. For example, both anomalously short \cite{poli2008} and anomalously long \cite{yang2010} spin relaxation times have been reported in superconducting aluminum.


In this paper, we summarize some of our recent experimental results on nonequilibrium charge and spin transport in nanoscale superconductors \cite{huebler2010,huebler2012b,wolf2013}, and perform additional numerical analysis to obtain more insight into the physical mechanisms.


\section{Results and Discussion}

\begin{figure}
\caption{(Color online) False color scanning electron miscroscopy image of one of our samples, together with the measurement scheme. The samples consist of a central superconducting wire (S), with normal-metal (N) and/or ferromagnetic (F) wires attached to it via tunnel contacts \cite{wolf2013}.}
\label{fig:sample}
\includegraphics[width=8.2cm,keepaspectratio]{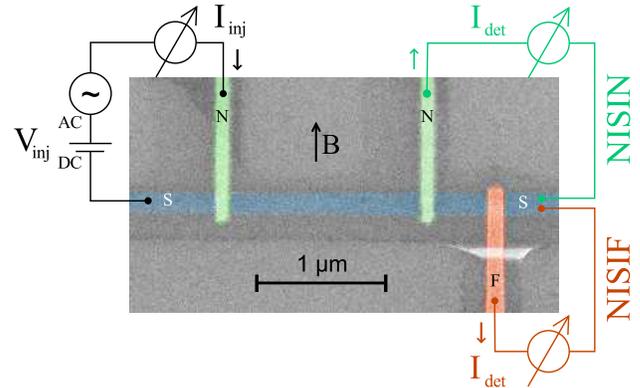}
\end{figure}

\Cref{fig:sample} shows a typical sample layout and measurement scheme. A central superconducting aluminum wire is contacted by several normal-metal (copper) or ferromagnetic (iron) electrodes attached via thin tunnel barriers. A dc bias voltage $V_\mathrm{inj}$ with a small superimposed low-frequency ac excitation is applied to one junction (injector), and the resulting current $I_\mathrm{inj}$ flowing into the junction is measured to determine the local differential conductance $g_\mathrm{loc}=dI_\mathrm{inj}/dV_\mathrm{inj}$. Simultaneously, the current $I_\mathrm{det}$ flowing out of a nearby detector junction is measured to obtain the nonlocal conductance $g_\mathrm{nl}=dI_\mathrm{det}/dV_\mathrm{inj}$. The nonlocal conductance was measured for different contact distances $d$, and different material combinations, where both injector and detector could be either normal (N) or ferromagnetic (F). These configurations will be labeled by $A$ISI$B$, where $A$ and $B$ denote the injector and detector contacts, respectively. Two examples (NISIN and NISIF) are indicated in \cref{fig:sample}. The measurements were carried out in a dilution refrigerator at temperatures down to about $50~\mathrm{mK}$, and with a magnetic field $B$ applied along the substrate plane parallel to the copper or iron wires. The thickness of the aluminum films was $t_\mathrm{Al}=12-30~\mathrm{nm}$, and for the thinnest films critical fields exceeding $2~\mathrm{T}$ were observed.

\begin{figure}
\caption{(a) Nonlocal conductance of one contact pair of an NISIN sample with $d=1~\mathrm{\mu m}$ as a function of injector bias $V_\mathrm{inj}$ for different magnetc fields $B$.
(b) Charge imbalance relaxation length $\lambda_{Q^*}$. Data taken from \cite{huebler2010}, lines are various model predictions explained in the text.}
\label{fig:lambdaQ}
\includegraphics[width=8.2cm,keepaspectratio]{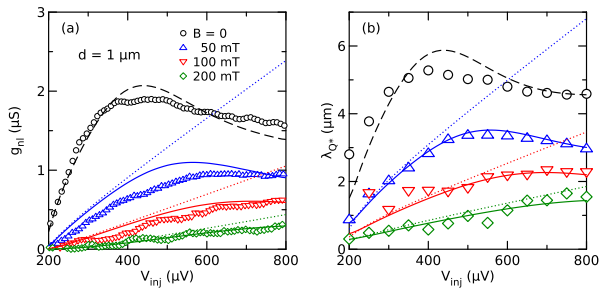}
\end{figure}

Before we discuss the spin signal observed using ferromagnetic detector junctions, we analyze the charge imbalance signal observed in an NISIN configuration. The aluminum film thickness of this sample was $t_\mathrm{Al}=30~\mathrm{nm}$, with a critical field $B_\mathrm{c}=0.53~\mathrm{T}$. Here, the effect of the applied field is mostly orbital pair breaking, and the Zeeman splitting of the density of states does not play a significant role. In \cref{fig:lambdaQ}(a), we show the nonlocal conductance $g_\mathrm{nl}$ of a pair of contacts at low temperature and for bias voltages above the energy gap $\Delta\approx 200~\mathrm{\mu eV}$ of the superconductor. From fitting $g_\mathrm{nl}$ at a given bias voltage for different contact distances to an exponential decay, we can obtain a bias-dependent charge relaxation length $\lambda_{Q^*}$ (see \cite{huebler2010} for details). The corresponding results are shown in \cref{fig:lambdaQ}(b).

Since we are interested here mostly in the behavior at finite magnetic fields, where Green's function methods are most appropriate, we model the data with the linearized kinetic equation derived by Schmid {\it et al.} \cite{schmid1975}. A simple analytical approximation neglecting cooling of the quasiparticles (see \cref{si:model}) yields the charge-imbalance relaxation length at low temperature
\begin{equation}
\lambda_\mathrm{Q^*} =  \xi\sqrt{\frac{N_1^2+N_2^2}{2N_2}}, \label{eqn:lambda_nocooling}
\end{equation}
where $N_1$ is the density of states in the superconductor, $N_2$ is a component of the anomalous Green's function, and $\xi$ is the dirty-limit coherence length. The nonlocal conductance due to charge imbalance within the same approximation is
\begin{equation}
g^\mathrm{CI}_\mathrm{nl} =G_\mathrm{inj}G_\mathrm{det}\int\frac{N_1^2}{N_1^2+N_2^2} \frac{\rho_\mathrm{N}\lambda_\mathrm{Q^*}}{2\mathcal{A}}  e^{-d/\lambda_\mathrm{Q^*}}f_0'(E-eV_\mathrm{inj}) dE,\label{eqn:gnl_nocooling}
\end{equation}
where $f_0'(E)$ is the derivative of the Fermi function.

In \cref{fig:lambdaQ}, we compare the model predictions to the experimental data. We proceed by first fitting $\lambda_{Q*}$ at finite magnetic fields with the simple ``no-cooling'' approximation \cref{eqn:lambda_nocooling}. Here, we assume that the pair-breaking strength follows the relation $\zeta=(B/B_\mathrm{c})^2/2$ for a magnetic field applied parallel to a thin film, and use the diffusion coefficient $D_\mathrm{N}$ as the single free fit parameter for all curves. These fits are shown as dotted lines in \cref{fig:lambdaQ}(b). As can be seen, a good fit can be made for the initial slope of the data, and we obtain $D_\mathrm{N}=70~\mathrm{cm}^2/\mathrm{s}$ from the fit, somewhat larger than the independent estimate ($40~\mathrm{cm}^2/\mathrm{s}$) from the resistivity. Without additional fitting, we can then plot the predictions for the nonlocal conductance according to \cref{eqn:gnl_nocooling} in \cref{fig:lambdaQ}(a). For large bias, the experimental data (both $g_\mathrm{nl}$ and $\lambda_{Q*}$) deviate downwards from the fits. Full numerical simulations including cooling, with the characteristic inelastic scattering time $\tau_E$ as the only remaining fit parameter, are shown as solid lines in \cref{fig:lambdaQ}. Excellent agreement with the experimental data for $\lambda_{Q*}$ can be achieved for $\tau_E=12~\mathrm{ns}$. The agreement for the nonlocal conductance is not as good as for $\lambda_{Q*}$, but still satisfactory. We finally attempted to fit the data at zero field, i.e., for $\zeta=0$. The predictions exceeded the experimental data by about a factor of two, both for  $g_\mathrm{nl}$ and $\lambda_{Q*}$ (not shown). We attribute this discrepancy to the fact that at zero applied field, any small additional source of pair breaking, such as gap anisotropy, magnetic impurities, spatial profile of the gap due to quasiparticle injection, etc., may contribute to charge relaxation \cite{lemberger1984b}. A reasonable fit (dashed lines) could be obtained by setting $\zeta=8\times10^{-4}$ to summarily account for all these pair-breaking perturbations. At zero field, we find a relaxation length of a few $\mathrm{\mu m}$, corresponding to characteristic time scales of a few ns. Recently, some experiments reported shorter time scales (sometimes by orders of magnitude) under similar conditions \cite{kleine2010,quay2013}. In contrast, our results are quantitatively consistent with the ``old'' knowledge obtained from experiments close to the critical temperature \cite{chi1979,stuivinga1983,mamin1984}, as well as more recent low-temperature experiments on the spatial decay of charge imbalance in thin wires \cite{yagi2006,arutyunov2011}. Both experimentally and theoretically, we find that the charge relaxation length decreases with increasing magnetic field, and is smallest at energies just above the gap. This is the parameter range where the spin signal is observed by ferromagnetic detectors described below. Also, in this parameter regime we can use the analytical ``no-cooling'' approximation \cref{eqn:gnl_nocooling} to describe charge imbalance.

\begin{figure}
\caption{Normalized nonlocal conductance of one contact pair in an FISIN (a) and NISIF (b) configuration as a function of $V_\mathrm{inj}$ for different magnetic fields $B$. Symbols are experimental data \cite{wolf2013}, lines are fits explained in the text.}
\label{fig:NISIF}
\includegraphics[width=8.2cm,keepaspectratio]{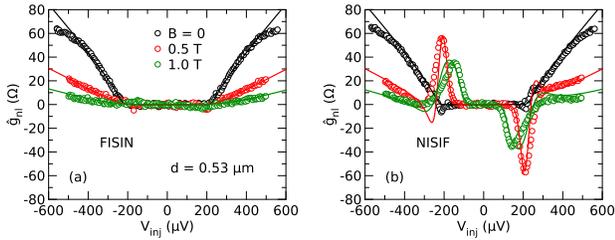}
\end{figure}

In \cref{fig:NISIF} we compare the nonlocal conductance for an FISIN (a) and NISIF (b) configuration, using the same pair of contacts, but reversing the roles of injector and detector. We plot here the normalized nonlocal conductance $\hat{g}_\mathrm{nl}=g_\mathrm{nl}/G_\mathrm{inj}G_\mathrm{inj}$, where $G_\mathrm{inj}$ and $G_\mathrm{inj}$ are the normal-state conductances of the injector and detector junctions, respectively. In the FISIN configuration, the nonlocal conductance is negligible at bias voltages below the gap. At bias voltages above the gap, the signal initially increases almost linearly, and then the slope decreases except for the highest magnetic fields. The signal is an even function of bias and can be attributed to charge imbalance, as described above, since the normal-metal detector is not sensitive to spin accumulation. The lines are fits to \cref{eqn:gnl_nocooling}. 

For the NISIF configuration, shown in \cref{fig:NISIF}(b), a similar signal is observed at $B=0$. Upon increasing the field, however, two additional peaks appear near the gap edge, with opposite sign for opposite bias polarity. These features can be attributed to spin injection into the Zeeman-split density of states of the superconductor \cite{giazotto2008,huebler2012b,wolf2013,quay2013}, which is probed by the ferromagnetic detector in this configuration. Spin-polarized tunneling can be described by two independent conductances $g_\downarrow$ and $g_\uparrow$ for the two spin orientations. The conductance is then given by the sum $g_\downarrow+g_\uparrow$, whereas the spin current is proportional to the difference $g_\downarrow-g_\uparrow$. The lines in \cref{fig:NISIF}(b) are the sum of the charge-imbalance contribution shown in \cref{fig:NISIF}(a) and an additional contribution  $g_{\textrm{nl}}^{S}\propto \left(g_\downarrow-g_\uparrow\right)$ to account for the spin signal. For the latter, we use parameters obtained from fits of the local conductance of the injector junction, leaving only the overall signal amplitude as a free fit parameter. As can be seen, a reasonable fit can be obtained over the entire bias range.

\begin{figure}
\caption{Charge relaxation length $\lambda_{Q^*}$ at a bias voltage of about $2\Delta$ (a) and spin diffusion length $\lambda_S$ (b) for  different samples as a function of normalized magnetic field $B/B_\mathrm{c}$. The samples have different number of ferromagnetic (F) and normal-metal (N) contacts, as indicated in the legend. Symbols are experimental data \cite{huebler2012b,wolf2013}, lines are fits explained in the text.}
\label{fig:lambdaS}
\includegraphics[width=8.2cm,keepaspectratio]{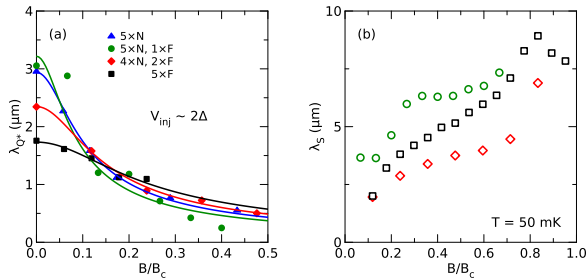}
\end{figure}

In \cref{fig:lambdaS}, we compare the charge and spin relaxation lengths of several samples with similar aluminum film properties as a function the normalized magnetic field $B/B_\mathrm{c}$. The samples have different numbers of ferromagnetic and normal-metal junctions, as indicated in the figure. In \cref{fig:lambdaS}(a), we plot the charge relaxation length $\lambda_{Q^*}$ obtained at a bias voltage of about $2\Delta$, where $\lambda_{Q^*}$ is usually largest at zero field (compare \cref{fig:lambdaQ}). $\lambda_{Q^*}$ is typically a few microns at zero field, and then quickly drops. The lines are fits to \cref{eqn:lambda_nocooling}. The spin relaxation length $\lambda_S$ is found by fitting the area $A$ of the spin-signal peaks as a function of contact distance to an exponential decay \cite{huebler2012b,wolf2013}. At small fields, $\lambda_S$ is similar to $\lambda_{Q^*}$, but then strongly increases with increasing field. At present, no theoretical model for high-field spin diffusion and relaxation in superconductors is available, therefore only a tentative interpretation is possible. The normal-state spin diffusion length in the samples is typically less than $500~\mathrm{nm}$, which means a tenfold increase in the  superconducting state. A possible relaxation mechanism could be a two-stage process of spin-flip scattering and recombination, which has been considered theoretically in a different context \cite{grimaldi1996,grimaldi1997}. A generalization of existing models for nonequilibrium transport in superconductors \cite{schmid1975,morten2004} to the case of large Zeeman splitting, treating both charge and spin degrees of freedom on an equal footing, would be highly desirable.

\begin{figure}
\caption{Spin relaxation length $\lambda_S$ (a) and amplitude $A$ of the spin signal (b) for  different samples as a function of temperature $T$. Symbols are experimental data \cite{huebler2012b,wolf2013}, lines are fits explained in the text.}
\label{fig:lambdaS_vs_T}
\includegraphics[width=8.2cm,keepaspectratio]{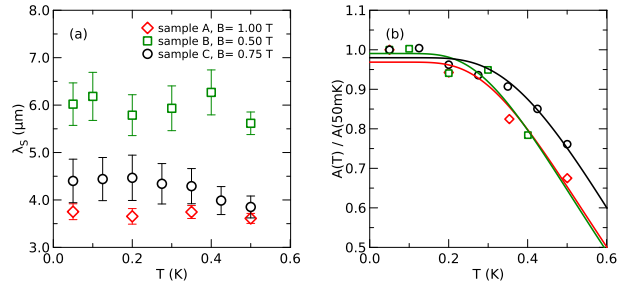}
\end{figure}

\Cref{fig:lambdaS_vs_T} shows the evolution of the spin relaxation length $\lambda_S$ and the amplitude $A$ of the spin signal as a function of temperature. $\lambda_S$ is independent of temperature within the accuracy of the experiment, similar to $\lambda_{Q^*}$ in the same temperature range \cite{huebler2010}.  In contrast, the signal amplitude decreases with increasing temperature. The spin injection rate proportional to $g_\downarrow-g_\uparrow$ inferred from the local conductance does not change appreciably in this temperature range, except for thermal broadening, which should not affect the overall peak area $A$. Thus, since neither injection nor relaxation cause the signal change, the decrease of signal amplitude must be related to the detection process. A simple model based on the tunnel Hamiltonian yields \cite{huebler2012b}
\begin{equation}
I_\mathrm{det} \propto S=\sum_\sigma \sigma \int N_{1\sigma}(E)\left[f_{\sigma}(E)-f_0(E)\right] dE,
\end{equation}
where $S$ is the net spin accumulation, $f_{\sigma}(E)$ is the quasiparticle distribution for spin $\sigma$ in the superconductor, and $f_0$ denotes the Fermi distribution in the ferromagnetic detector junction. As can be seen, the detector signal is proportional to the difference of the distribution functions in the superconductor and ferromagnet. The former is determined by spin injection, whereas the latter can be assumed to be (nearly) at equilibrium at the bath temperature. Therefore, we can expect the spin signal to decrease as the bath temperature is raised. A very simple model to describe this drop can be obtained by assuming that nonequilibrium injection raises the effective temperature of the quasiparticles inside the superconductor to about $1~\mathrm{K}$, as we have found in similar structures with normal-metal junctions \cite{kolenda2013}, and that most quasiparticles have an energy close to the energy gap $E_\mathrm{g}$, which is typically around $0.5-0.75\times \Delta_0$ at the fields of the experiments. Then, the spin signal should be proportional to $f_0(E_\mathrm{g},1~\mathrm{K})-f_0(E_\mathrm{g},T)$. Fits to this model are shown in \cref{fig:lambdaS_vs_T}(b). As can be seen, the agreement is quite good, despite the oversimplification of the model. We note that usually the current through an NIS junction does not depend on the temperature of the normal metal due to particle-hole symmetry. This is no longer true if a spin-dependent density of states in the superconductor is combined with a spin-dependent tunnel conductance, as it is the case in our experiment. For this case, large thermoelectric effects driven by the temperature difference between superconductor and ferromagnet have been predicted recently \cite{machon2013,ozaeta2013}.

\section{Conclusions}

We have presented an analysis of our recent experiments on spin and charge transport in nanoscale superconductors at very low temperatures and high magnetic fields. We find that charge imbalance can be described surprisingly well with existing models, despite the fact that they were initially developed for experiments close to the critical temperature. Charge relaxation is very fast at energies just above the gap. This is the bias regime where we observe long-range spin transport in the presence of a Zeeman splitting of the density of states. By comparing the relaxation lengths for charge and spin, we can conclude that spin currents in this regime are nearly chargeless. While no detailed model of spin transport and relaxation is available yet, we find that simple models based on the tunnel Hamiltonian explain the dependence of spin injection and detection on bias, magnetic field and temperature. The ability to create and transport pure spin currents in superconductors may be useful for future superconducting spintronics devices. Further, our analysis of the temperature dependence hints at the importance of new thermoelectric effects in nanoscale superconductor-ferromagnet hybrids.

\begin{suppinfo}
\sifile{beckmann_S1.pdf}{PDF}{Model}\label{si:model}
\end{suppinfo}
\begin{acknowledgements}
We acknowledge financial support by DFG grant BE-4422/1-1 and the competence network ``Functional Nanostructures'' of the Baden-W\"urttemberg-Stiftung, and W. Belzig and M. Eschrig for stimulating discussions.
\end{acknowledgements}

\bibliography{beckmann}



\end{document}


\title{Charge and spin transport in mesoscopic superconductors - Supplementary Material S1}
\author{M. J. Wolf, F. H\"ubler, S. Kolenda, and D. Beckmann\\Karlsruher Institut f\"ur Technologie (KIT), Institut f\"ur Nanotechnologie}
\date{November 28, 2013}
\maketitle

\section{Spectral properties}

To obtain the spectral properties of the superconductor needed to describe our experiments, we solve the Usadel equation \cite{usadel1970,belzig1999}, including all the perturbations needed for our purpose. Pair breaking is included by the dimensionless pair-breaking strength 
\begin{equation}
\zeta=\frac{1}{2}\left(\frac{B}{B_\mathrm{c}}\right)^2
\end{equation}
for a thin film with a magnetic field $B$ applied parallel to the film plane, with critical field $B_\mathrm{c}$. It is related to the magnetic-impurity scattering time $\tau_\mathrm{s}$ used in \cite{schmid1975} by $\zeta=\hbar/2\Delta_0\tau_\mathrm{s}$, where $\Delta_0$ is the superconducting pair potential at $T=0$ and $B=0$. The quasiparticle life time due to electron-phonon scattering is included by adding a small imaginary part (Dynes parameter \cite{dynes1978}) $\Gamma=\hbar/2\tau_E$ to the energy, where $\tau_E$ is the inelastic scattering time (see below). We also include the Zeeman energy $\pm\mu_\mathrm{B}B$, and spin-orbit scattering with scattering strength $b_\mathrm{so}=\hbar/3\Delta \tau_\mathrm{so}$. Since all our junctions have tunnel barriers, we neglect the proximity effect, and assume that the spectral properties of our superconducting wires are independent of position, i.e., we neglect gradient terms in the Usadel equation. The complete Usadel equation with all terms included can be found, e.g., in \cite{alexander1985}.

\section{Kinetic equation for charge imbalance}
We describe charge imbalance using the kinetic equation derived by Schmid {\it et al.} \cite{schmid1975}. To apply the kinetic equation to our experiments, we consider a superconducting wire of cross-section $\mathcal{A}$ and normal-state diffusion coefficient $D_\mathrm{N}$ along the $x$-axis, an injector junction with normal-state conductance $G_\mathrm{inj}$ placed at $x=0$, and a detector junction with normal-state conductance $G_\mathrm{det}$ placed at $x=d$. We assume that both the wire cross-section and the junction width are small compared to the charge-imbalance relaxation length. We are interested only in the stationary case.  The kinetic equation for the transverse-mode distribution function $f_\mathrm{T}$ then reads
\begin{equation}
D_\mathrm{N}M_\mathrm{T}(E,E)\frac{d^2}{dx^2} f_\mathrm{T}(E) + K_\mathrm{T}(E,\{f_\mathrm{T}\}) + Q_\mathrm{T}(E) + P_\mathrm{T}(E) =0\label{eqn:kineq}
\end{equation}
Here, $M_\mathrm{T}(E,E') = N_1(E) N_1(E')+N_2(E) N_2(E')$, and $N_1$ and $N_2$ are the real parts of the normal and anomalous Green's function found solving the Usadel equation ($N_1$ is the density of states). The collision integral describing energy relaxation is
\begin{eqnarray*}
K_\mathrm{T}(E,\{f_\mathrm{T}\})& = & - \int dE'\frac{\mu(E-E')M_\mathrm{T}(E,E')}{\cosh\left(E/2k_\mathrm{B}T\right)\cosh\left(E'/2k_\mathrm{B}T\right)\sinh\left((E-E')/2k_\mathrm{B}T\right)} \\
 & \times & \left[\cosh^2\left(\frac{E}{2k_\mathrm{B}T}\right)f_\mathrm{T}(E)-\cosh^2\left(\frac{E'}{2k_\mathrm{B}T}\right)f_\mathrm{T}(E')\right]
\end{eqnarray*}
where
\begin{eqnarray*}
\mu(E) & = & \frac{\mathrm{sign}(E) E^2}{14\zeta(3)(k_\mathrm{B}T_\mathrm{c})^3\tau_E}
\end{eqnarray*}
within the Debye model, and $\tau_E$ is the inelastic scattering time for electrons at the Fermi surface in the normal state at $T_\mathrm{c}$.
Charge relaxation is given by
\begin{eqnarray*}
Q_\mathrm{T}(E)    & = & -2\frac{\Delta}{\hbar} N_2(E)f_\mathrm{T}(E),
\end{eqnarray*}
and the injection rate is
\begin{eqnarray*}
P_\mathrm{T}(E)    & = & \frac{G_\mathrm{inj}}{2N_0 \Omega e^2} N_1(E) f_\mathrm{inj}(E,V)  \\
f_\mathrm{inj}(E,V)  & = & \frac{1}{4}\left[\tanh\left(\frac{E+eV}{2k_\mathrm{B}T}\right)-\tanh\left(\frac{E-eV}{2k_\mathrm{B}T}\right)\right]
\end{eqnarray*}
where $\Omega$ is the injection volume and $N_0$ is the density of states of the superconductor per spin. The current flowing from S to N through the detector junction held at zero bias voltage is given by
\begin{equation}
I_\mathrm{det} = \frac{G_\mathrm{det}}{e} \int dE N_1(E) f_\mathrm{T}(E).\label{eqn:idet_general}
\end{equation}

\subsection{Approximate analytical solution (without cooling)}
At low temperatures, inelastic scattering is expected to freeze out. To obtain a simple analytical solution for $T\rightarrow 0$, we therefore neglect the collision integral, i.e., the cooling of the quasiparticles. We also assume that the injector junction is infinitesimally small, with inverse injection volume $\Omega^{-1}=\mathcal{A}^{-1}\delta(x)$. Then the kinetic equation can be easily solved by inserting the Ansatz
\begin{equation}
f_\mathrm{T}(x,E) = a(E)e^{-|x|/\lambda_\mathrm{Q^*}(E)},
\end{equation}
which yields the two conditions
\begin{eqnarray}
\lambda_\mathrm{Q^*}(E) & = &  \xi\sqrt{\frac{N_1^2+N_2^2}{2N_2}} \label{eqn:lambda_nocooling} \\
a(E) & = &  G_\mathrm{inj}\frac{\rho_\mathrm{N}\lambda_\mathrm{Q^*}}{2\mathcal{A}} \frac{N_1}{N_1^2+N_2^2} f_\mathrm{inj}(E,V)
\end{eqnarray}
where we have introduced the dirty-limit coherence length $\xi=\sqrt{\hbar D_\mathrm{N}/\Delta}$ and the normal-state resistivity of the superconductor $\rho_\mathrm{N}=(2N_0e^2D_\mathrm{N})^{-1}$. 
Inserting this solution into (\ref{eqn:idet_general}) yields the detector current
\begin{equation}
I_\mathrm{det} = \frac{G_\mathrm{inj}G_\mathrm{det}}{e}\int dE \frac{\rho_\mathrm{N}\lambda_\mathrm{Q^*}}{2\mathcal{A}} \frac{N_1^2}{N_1^2+N_2^2} e^{-d/\lambda_\mathrm{Q^*}} f_\mathrm{inj}(E,V)
\end{equation}
Using the symmetry properties of the various quantities, we finally obtain
\begin{equation}
g_\mathrm{nl}=\frac{dI_\mathrm{det}}{dV_\mathrm{inj}} =G_\mathrm{inj}G_\mathrm{det}\int\frac{N_1^2}{N_1^2+N_2^2} \frac{\rho_\mathrm{N}\lambda_\mathrm{Q^*}}{2\mathcal{A}}  e^{-d/\lambda_\mathrm{Q^*}}f'(E-eV_\mathrm{inj}) dE,\label{eqn:gnl_nocooling}
\end{equation}
where $f'(E)=\cosh^{-2}\left(E/2k_\mathrm{B}T\right)/4k_\mathrm{B}T$ is the derivative of the Fermi function.

\subsection{Numerical solution}
For a full numerical solution, we discretize the equation on grid points $E_i$ and $x_k$. The kinetic equation (\ref{eqn:kineq}) then turns into a linear equation system for the distribution function $f_{ik}=f_\mathrm{T}(x_k,E_i)$.  This equation system is solved by standard library routines \cite{lapack}. The size of the grid points is chosen small enough to not affect the results ($\delta E\approx 10~\mathrm{\mu eV}$, $\delta x=500~\mathrm{nm}$).  From the numerical solution, we obtain the nonlocal conductance as a function of bias and contact distance, and analyze it in the same way as the experimental data \cite{huebler2010} to obtain the relaxation length.

\section{Spin imbalance}

In order to describe the local conductance of the injector junctions, we use the theory of tunneling in superconductors in high magnetic field \cite{maki1964b,meservey1994}. The contribution of a single spin projection $\sigma=\pm 1$ to the tunnel conductance is given by
\begin{equation}
  g_\sigma = \frac{G_\mathrm{inj}}{2}\left(1-\sigma P_\mathrm{inj}\right) \int N_{1\sigma}(E) f' dE,
  \label{eqn_gsigma}
\end{equation}
where $P_\mathrm{inj}$ is the spin polarization of the tunnel conductance, and $N_{1\sigma}(E)$ is the density of states in the superconductor for spin projection $\sigma$, obtained by solving the Usadel equation. The injector conductance is given by the sum of the two spin contributions, 
\begin{equation}
 g_\mathrm{loc}=g_\downarrow+g_\uparrow,\label{eqn_gloc}
\end{equation}
whereas the differential spin current
\begin{equation}
\frac{dI_\sigma}{dV_\mathrm{inj}}\propto g_\downarrow-g_\uparrow\label{eqn_spin_injection}
\end{equation}
is proportional to their difference. From fits of the local conductance, we can therefore infer the bias-dependent spin injection rate. A simple tunnel Hamiltionian model for the spin-related contribution to the detector current yields \cite{zhao1995,huebler2012b}
\begin{equation}
I_\mathrm{det}^S = \frac{G_\mathrm{det}P_\mathrm{det}}{2e} \sum_\sigma \sigma \int N_{1\sigma}(E)\left[f_{\sigma}(E)-f_0(E)\right] dE,\label{eqn_idet}
\end{equation}
where $f_{\sigma}(E)$ is the quasiparticle distribution for spin $\sigma$ in the superconductor, and $f_0$ denotes the Fermi distribution in the ferromagnetic detector junction. Combining (\ref{eqn_spin_injection}) and (\ref{eqn_idet}), we expect the contribution of spin accumulation to the nonlocal conductance to be
\begin{equation}
g_{\textrm{nl}}^{S}\propto P_\mathrm{det}\left(g_\downarrow-g_\uparrow\right).
\end{equation}

\bibliography{beckmann_S1.bib}
